\documentclass[preprint,showpacs]{revtex4}

\usepackage[latin1]{inputenc}
\usepackage{graphicx}%
\usepackage{dcolumn}
\usepackage{amsmath}

\makeatletter
\def\btt#1{\texttt{\@backslashchar#1}}%
\DeclareRobustCommand\bblash{\btt{\@backslashchar}}%
\makeatother

\begin{document}

 \title{Traffic Network Optimum    Principle  --  Minimum Probability of Congestion Occurrence  }

\mark{Traffic Network Optimum   }

\author{Boris S. Kerner $^1$ }

\address{$^1$ 
Daimler AG, GR/PTF, HPC:  G021, 71059 Sindelfingen, Germany 
}

%\date{October 10, 2010}

\pacs{89.40.-a, 47.54.-r, 64.60.Cn, 05.65.+b}

\begin{abstract}
We introduce an optimum principle for  a    vehicular traffic network with   road bottlenecks. This 
 network  {\it breakdown minimization (BM) principle} states
that the network  optimum      is reached, when  link   flow rates   
are assigned  in the network in such a way that the  
probability 
for spontaneous occurrence of
 traffic breakdown     at one of the network bottlenecks during a given observation time  
   reaches the minimum possible value.  
 Based on numerical simulations with a stochastic three-phase traffic flow model, we show that
in comparison to 
  the well-known Wardrop's    principles    
   the application of the BM principle permits considerably greater network inflow rates    at which {\it no} traffic breakdown occurs and, therefore,
free flow remains in the whole network.
     \end{abstract}

\maketitle

\section{Introduction}
 \label{Int}

Under   small enough network inflow rates,  drivers move at their desired (or permitted) speeds. 
Usually,  there are several   alternative  routes from an origin to a destination in a  network for which
 travel times are different but close to each other.   When network inflow rates  increase considerably, 
traffic congestion occurs due to traffic breakdown
causing a sharply increase in the route travel times. 
Thus one of the  theoretical problems of traffic networks is to find
an {\it optimal} feedback
dynamic traffic network assignment between alternative  routes
 that prevents traffic breakdown under great enough network inflow rates 
 while maintaining free flow in the  network (see, e.g., review~\cite{Rakha2008}). 
 Traffic breakdown occurs mostly at a  bottleneck  and leads to the emergence of spatiotemporal congested
traffic patterns.
The bottleneck can result from on- and off-ramps, a road gradient, etc.

\begin{figure}
\begin{center}
\includegraphics*[width=14 cm]{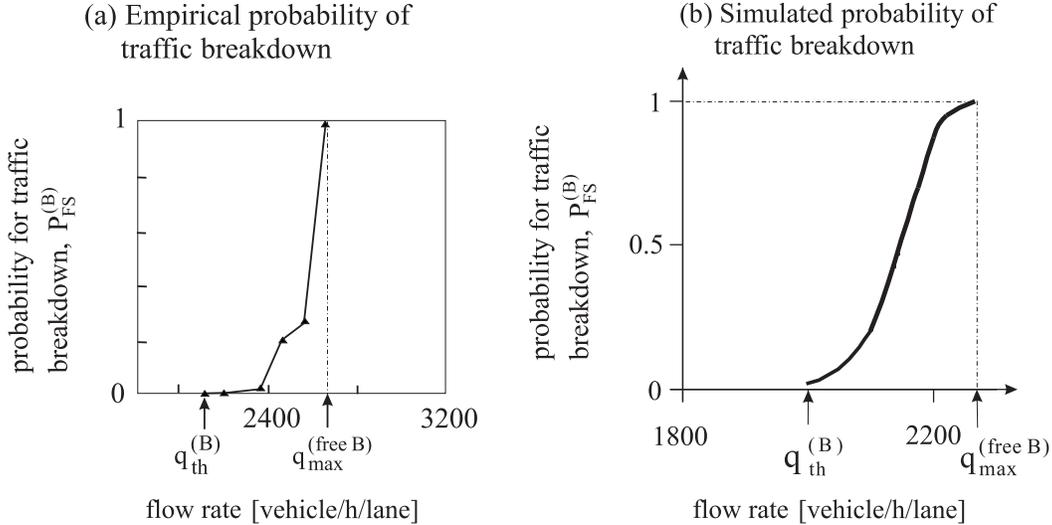}
\caption{Empirical (a) and simulated (b) probability of traffic breakdown
at an on-ramp bottleneck. Fig. (a) is taken from Persaud {\it et al.}~\cite{Persaud}.
Fig. (b) is taken from Fig.~4.2 of~\cite{KernerBook2}. In (a) the averaging time interval for traffic variables $T_{\rm av}=$ 10 min~\cite{TOB}.
In (b) the observation time of traffic flow $T_{\rm ob}=$ 15 min.
\label{Probability} } 
\end{center}
\end{figure}

An empirical feature of traffic breakdown at a bottleneck is as follows~\cite{KernerBook,KernerBook2}.
   Traffic breakdown is a local {\it first-order}
   phase transition from free flow to synchronized flow (F$\rightarrow$S transition).  The    feature  
has  been explained  in three-phase traffic theory~\cite{KernerBook}
 in which there are three phases: 1. Free flow (F). 2. Synchronized flow (S). 3. Wide moving jam (J).  
 Synchronized flow and wide moving jam are associated with
 congested traffic.  
A wide moving jam   exhibits the characteristic jam feature   to propagate through bottlenecks
  while  maintaining the mean velocity of the downstream jam front.
  In contrast, the downstream front of synchronized flow
is often   fixed  at the bottleneck. It has been found that 
there is a broad range $q^{\rm (B)}_{\rm th}\leq q\leq q^{\rm (free \  B)}_{\rm max}$~\cite{FlowMaxMin} of the link (arc) flow rate 
$q$ 
within which free flow at the bottleneck is in a metastable state 
with respect to traffic breakdown (Fig.~\ref{Probability}). 
The greater the  flow rate $q$ in comparison with  $q^{\rm (B)}_{\rm th}$,
the smaller   the critical amplitude of a disturbance in free flow  whose growth leads to the breakdown, i.e., the 
greater the  probability $P^{\rm (B)}_{\rm FS}$ of the breakdown occurrence during a given observation time $T_{\rm ob}$. 
At $q<q^{\rm (B)}_{\rm th}$ probability $P^{\rm (B)}_{\rm FS}=$ 0, i.e., no traffic breakdown   occurs,
 while at the maximum flow rate $q=q^{\rm (free \  B)}_{\rm max}$
traffic breakdown occurs already due to a small disturbance, i.e., with probability $P^{\rm (B)}_{\rm FS}=$ 1. 

Most   network optimization theories (see e.g.,~\cite{Rakha2008,Dif,Wahle,Davis,Davis1})  
 are based on the application of      user  equilibrium  (UE) 
and   system optimum (SO) principles introduced by Wardrop~\cite{Wardrop}:
(i)  {\it Wardrop's UE principle:}  traffic on a network distributes itself in such a way that the travel times on 
all routes used from any origin to any destination are equal, while all unused routes have equal or greater 
travel   times. 
(ii) {\it Wardrop's SO principle:}   the network-wide travel time should be a minimum. 
The Wardrop's    principles reflect either the wish of  drivers
to reach their destinations as soon as possible (UE) or the wish of network operators to reach the minimum 
network-wide travel time   (SO). 

However,     the Wardrop's    principles 
do  {\it not}
take into account that with some probability traffic breakdown    occurs in the network,  
when   the link flow rate   for one of the network bottlenecks exceeds
 $q^{\rm (B)}_{\rm th}$. This breakdown    leads usually  to  
   spatiotemporal      congestion propagation~\cite{KernerBook,KernerBook2}. 
 Such    congestion growth within the network causes the associated growth of  
 link travel times; as a result, under congestion conditions as has been shown
 by Wahle and Schreckenberg with colleagues~\cite{Wahle} and Davis~\cite{Davis,Davis1,Davis2} usually   no true Wardrop's equilibrium can be found.

 In this article, we introduce a network  {\it breakdown minimization (BM) principle}
   based on the empirical features of traffic breakdown. The application of the BM principle 
  should    minimize   probability of congestion occurrence 
  in the whole network.    
  We show that the BM principle   leads to  considerably greater network inflow rates 
     at which free flows  remain in the 
   network than under application of the Wardrop's SO and UE principles. 
   
\section{Network   breakdown minimization (BM) principle \label{S_BM}}

The    BM principle is as follows: 
 \begin{itemize}  
\item  The optimum   of a   traffic   network    with $M$ links and $N$   bottlenecks  
   is reached,  when  link   inflow rates   
are assigned  in the network in such a way that the  
probability 
 \begin{equation}
P^{\rm (N)}_{\rm FS, net}=  1- \prod^{\rm N}_{k=1} (1-P^{\rm (B,{\it k})}_{\rm FS})
\label{3Phase}
\end{equation}
for spontaneous occurrence of
 traffic breakdown at one of the network bottlenecks  during a given observation time $T_{\rm ob}$  
   reaches the minimum possible value,
  i.e., the network optimum is reached at
 \begin{equation}
\min_{q_{1},q_{2},...,q_{\rm M}} \{P^{\rm (N)}_{\rm FS, net}(q_{1},q_{2},...,q_{\rm M})\}. 
\label{3Phase2}
\end{equation}
In (\ref{3Phase}), (\ref{3Phase2}), $q_{m}$ is the link inflow rate
for a link with index $m$; $m=1,2,..., M$, where $M>1$;
$k= 1,2,..., N$ is bottleneck 
index~\footnote{Note that even if a 
network link is a spatially homogeneous road (i.e., the road without on- and off-ramps,
without road gradients, curves, etc.), nevertheless within a flow rate range
$q_{\rm th}\leq q \leq q^{\rm (free)}_{\rm max}$ traffic breakdown  occurs 
on this link spontaneously with   probability $P_{\rm FS}>$ 0, however, at a random location
within the link~\cite{KernerBook}. In (\ref{3Phase}), this link can also be considered containing a
  bottleneck with the threshold flow rate $q^{\rm (B)}_{\rm th}=q_{\rm th}$ 
  and maximum flow rate $q^{\rm (free \  B)}_{\rm max}=q^{\rm (free)}_{\rm max}$.}, 
$N> 1$;
$P^{\rm (B,{\it k})}_{\rm FS}$ is probability that during 
the time interval $T_{\rm ob}$ traffic breakdown occurs at  bottleneck   $k$.
The BM principle (\ref{3Phase2}) can be applied as long as       free flow conditions remain in the network.
In general, the BM principle (\ref{3Phase2}) is devoted to the optimization of large, complex vehicular traffic networks  
consisting
of a great number of links $M\gg 1$.
\end{itemize} 
  The BM principle (\ref{3Phase2}) is  equivalent to 
  \begin{equation}
\max_{q_{1},q_{2},...,q_{\rm M}} \{P^{\rm (N)}_{\rm C, net}(q_{1},q_{2},...,q_{\rm M})\},
\label{3Phase3}
\end{equation}
 where 
  \begin{equation}
P^{\rm (N)}_{\rm C, net}=   \prod^{\rm N}_{k=1} P^{\rm (B,{\it k})}_{\rm C} 
\label{3Phase4}
\end{equation}  
is the probability that 
 during     time interval $T_{\rm ob}$  free flows remain
   in the  network,  i.e.,   that
traffic breakdown occurs at {\it none} of the  bottlenecks;
$P^{\rm (B, {\it k})}_{\rm C}=1-P^{\rm (B, {\it k})}_{\rm FS}$. 

For a complete formulation of the
optimization principle (\ref{3Phase2}) (or  (\ref{3Phase3})), link flow rates $q_{\rm m}$  should  be   connected
with network inflow rates. To reach this goal in a general case of a dynamic
traffic assignment in a traffic network, one should use
a dynamic traffic flow model~\cite{Rakha2008}. This dynamic model
should calculate spatiotemporal dynamics
of vehicular  traffic variables within the network under
given network inflow rates that can be time-functions.

However, for the simplicity of simulations of the BM principle 
(\ref{3Phase2}) discussed below in Sec.~\ref{Sim}, we will use
here a static traffic assignment for which
 the following well-known    constraints are applied~\cite{Rakha2008}:
\begin{equation}
 \sum_{i}{\varphi^{rw}_{i}} =q_{rw} \quad \forall \ r,w,  
 \end{equation}
 \begin{equation}
 \varphi^{rw}_{i}\geq 0 \quad \forall \ i,r,w,
 \end{equation}
 \begin{equation}
  q_{m}=\sum_{r}{\sum_{w}{\sum_{i}{\varphi^{rw}_{i} \delta^{rw}_{m, i} }}} \quad \forall \ m,
  \end{equation} 
where $ q_{rw}$ is the total flow rate of vehicles going from origin $r$ to destination $w$;
 $\varphi^{rw}_{i}$ is the flow rate of   vehicles going from $r$ to   $w$ on   route (path) $i$; 
 \begin{eqnarray}
\delta^{r s}_{m, i}= \left\{
\begin{array}{ll}
1 &  \textrm{if link $m$ is on route $i$} \\
0 &  \textrm{otherwise}.
\end{array} \right.
\end{eqnarray}

 \section{Simulations: Comparison of the BM     and Wardrop's    principles}
 \label{Sim}
 
 \subsection{Model}

\begin{figure}
\begin{center}
\includegraphics*[width=14 cm]{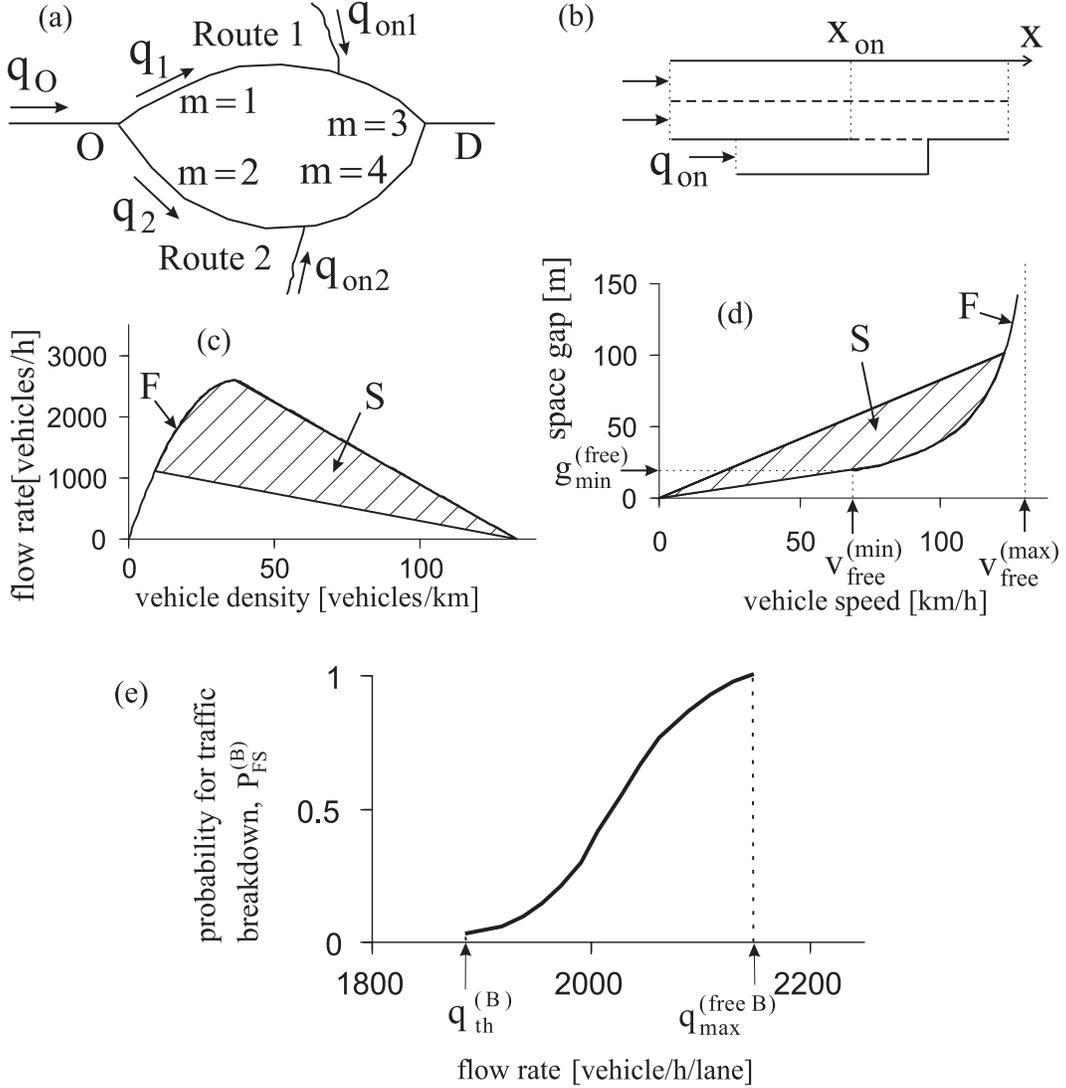}
\caption{Explanation of model: (a, b) Sketch of a simple network with two routes 1, 2 (a) and  a route model        (b);
bottleneck parameters are the same as those in~\cite{KKl2003A,KKl2009,KKl2010}. (c, d) Model steady states
in the flow--density  (c) and space-gap--speed planes (d); F -- free flow, S -- synchronized flow.
(e) Probability of spontaneous traffic breakdown at on-ramp bottleneck
as function of the flow rate  downstream of the bottleneck at $q_{\rm on}=$ 1000 vehicles/h
for $T_{\rm ob}=$ 40 min.
\label{Network_Model} } 
\end{center}
\end{figure}

We   compare  the BM (\ref{3Phase2})   and Wardrop's    principles  
through their application   for a  simple network  with only two alternative routes  1 and 2 with lengths
 $L_{1}$ and $L_{2}$ (with $L_{2}>L_{1}$)
 for vehicles moving from origin $O$ to destination $D$
  (Fig.~\ref{Network_Model} (a))
used often for studies of   traffic control
with Wardrop's   principles~\cite{Wahle,Davis,Davis1,Davis2}.  

In our model, we assume that    routes 1 and 2 are    two-lane roads with   on-ramp bottlenecks   (Fig.~\ref{Network_Model} (b))
whose
 on-ramp inflow rates $q_{\rm on1}$ and 
  $q_{\rm on  2}$ are given constants.
 Thus    the network optimization is performed only through the assignment of a network inflow with the rate $q_{\rm O}$    
 between links $m=1, 2$  
   on routes  $i=1, 2$ (Fig.~\ref{Network_Model} (a)). We designate link  flow rates  and   travel times, respectively, as follows:
  for links $m=1,3$ on route 1 by
  $q_{\rm 1}$, $q_{\rm 3}$  and $T_{1,1}$, $T_{3,1}$; for links $m=2,4$ on route 2 by
  $q_{2}$, $q_{4}$   and  $T_{2,2}$, $T_{4,2}$ (Fig.~\ref{Network_Model} (a)), where
 $q_{3}=q_{1}+q_{\rm on  1}$,
  $q_{4}=q_{2}+q_{\rm on  2}$. 
  Travel times on routes 1 and 2 are   
  $T_{1}=T_{1,1}+T_{3,1}$ and $T_{2}=T_{2,2}+T_{4,2}$, respectively.
The BM principle (\ref{3Phase2}) as well as Wardrop's  UE and SO  principles can be written respectively as
  follows:
   \begin{eqnarray}
   \label{3Phase2ex}
{\rm BM:} \quad \min_{q_{1},q_{2}} \{ 1- (1-P^{\rm (B, 1)}_{\rm FS}(q_{1}+q_{\rm on1}))(1- 
\\ - P^{\rm (B,  2)}_{\rm FS}(q_{2}+q_{\rm on2}))\}, \ \nonumber
  q_{1}+q_{2}=q_{\rm O}, 
 \end{eqnarray}
    \begin{equation}
{\rm UE:} \quad  T_{1}(q_{1},q_{\rm on1})=  T_{2}(q_{2},q_{\rm on2}), \ q_{1}+q_{2}=q_{\rm O},
  \label{UE_Eq}
  \end{equation}
   \begin{eqnarray}
     \label{SO_Eq}
{\rm SO:} \quad  \min_{q_{1}, q_{2}} \{q_{1}T_{1,1}+(q_{1}+q_{\rm on1})T_{3,1} +q_{2}T_{2,2}+   \\  + (q_{2}+q_{\rm on2})T_{4,2}\}, \ \nonumber
  q_{1}+q_{2}=q_{\rm O}.
  \end{eqnarray}
  
  Travel times
  $T_{1,1}$, $T_{3,1}$, $T_{2,2}$, $T_{4,2}$   are found via probe
 vehicles leaving the related links. These travel times are used in
   the UE    (\ref{UE_Eq}) and  SO  (\ref{SO_Eq}) principles for calculations of  $q_{1}$,  $q_{2}$
   as long as the  probe vehicles have moved
 in  {\it free flows}; this explains why only the associated time intervals are shown in related figures below~\cite{Timelag}.
  
  For simulations, we use a discrete  version~\cite{KKl2009}  
  of the Kerner-Klenov stochastic three-phase traffic flow model of~\cite{KKl2003A} that   shows
  the empirical features of traffic breakdown including the resulting flow-dependence of breakdown probability  
  $P^{\rm (B)}_{\rm FS}$ (Fig.~\ref{Network_Model} (e))
  used in  (\ref{3Phase2ex})~\cite{GM_model}. The model reads as follows:
  \begin{equation}
v_{n+1}=\max(0, \min(v_{{\rm free},n}, \tilde v_{n+1}+\xi_{n}, v_{n}+a \tau, v_{{\rm s},n} )),
\label{final}
\end{equation}
\begin{equation}
\label{next_x}
x_{n+1}= x_{n}+v_{n+1}\tau,
\end{equation}
where 
 $n=0, 1, 2, ...$ is number of time steps,
$\tau$ is a time step, $x_{n}$ and $v_{n}$ are the vehicle coordinate and speed
at time step $n$,   $a$ is the maximum acceleration,
$\tilde v_{n}$ is the vehicle speed  without  speed fluctuations $\xi_{n}$,
$v_{{\rm s}, n}$ is a safe speed. 

The physics of this model as well as initial and boundary conditions
used in  simulations have already been considered in detail in Sec.~16.3 of the book~\cite{KernerBook}.
In   accordance with the fundamental hypothesis of three-phase traffic theory~\cite{KernerBook,KernerBook2},
 steady states of synchronized flow cover a 2D-region in the flow--density plane (Fig.~\ref{Network_Model} (c)).
Speed fluctuations $\xi_{n}$, functions  $\tilde v_{n}$, $v_{{\rm s}, n}$,
 rules for lane changing 
and model parameters used here   are taken from~\cite{KKl2010} (see Appendix~\ref{CA_Ap}). The one exception from the model version of~\cite{KKl2010}  
 is that a free  flow speed $v_{{\rm free},n}$ rather than to be a constant
  depends
on space gap $g_{n}$ to the preceding vehicle: 
\begin{equation}
v_{{\rm free}, n}=v_{\rm free}(g_{n}),
\end{equation}
where
\begin{equation}
v_{\rm free}(g)=\max [v_{\rm free}^{\rm (max)}(1-\kappa d/(g+d)), v_{\rm free}^{\rm (min)}],
\end{equation}
$\kappa$, $v_{\rm free}^{\rm (max)}$ are  given constants,  $v_{\rm free}^{\rm (min)}$ (Fig.~\ref{Network_Model} (d)) is   constant      
  found from the equations
  \begin{equation}
v^{\rm (free)}_{\rm min}= g^{\rm (free)}_{\rm min}/\tau,
\end{equation} 
  \begin{equation}
v^{\rm (free)}_{\rm min}=v_{\rm free}(g^{\rm (free)}_{\rm min}).
\end{equation}

\subsection{Critical flow rate   for traffic breakdown}

   In simulations,  we study  the spontaneous occurrence of traffic breakdown
  at one of the bottlenecks   in the 
 network
  (Fig.~\ref{Network_Model} (a))    during a given observation time  $T_{\rm ob}=$ 40 min
 (where $T_{\rm ob}>T_{1}, \ T_{2}$) at given on-ramp inflow rates   $q_{\rm on  1}$, $q_{\rm on  2}$
 under  network optimization based     on  the application of each of the
principles   (\ref{3Phase2ex}),  (\ref{UE_Eq}), and   (\ref{SO_Eq}). 

  \begin{figure}
\begin{center}
\includegraphics*[width=13 cm]{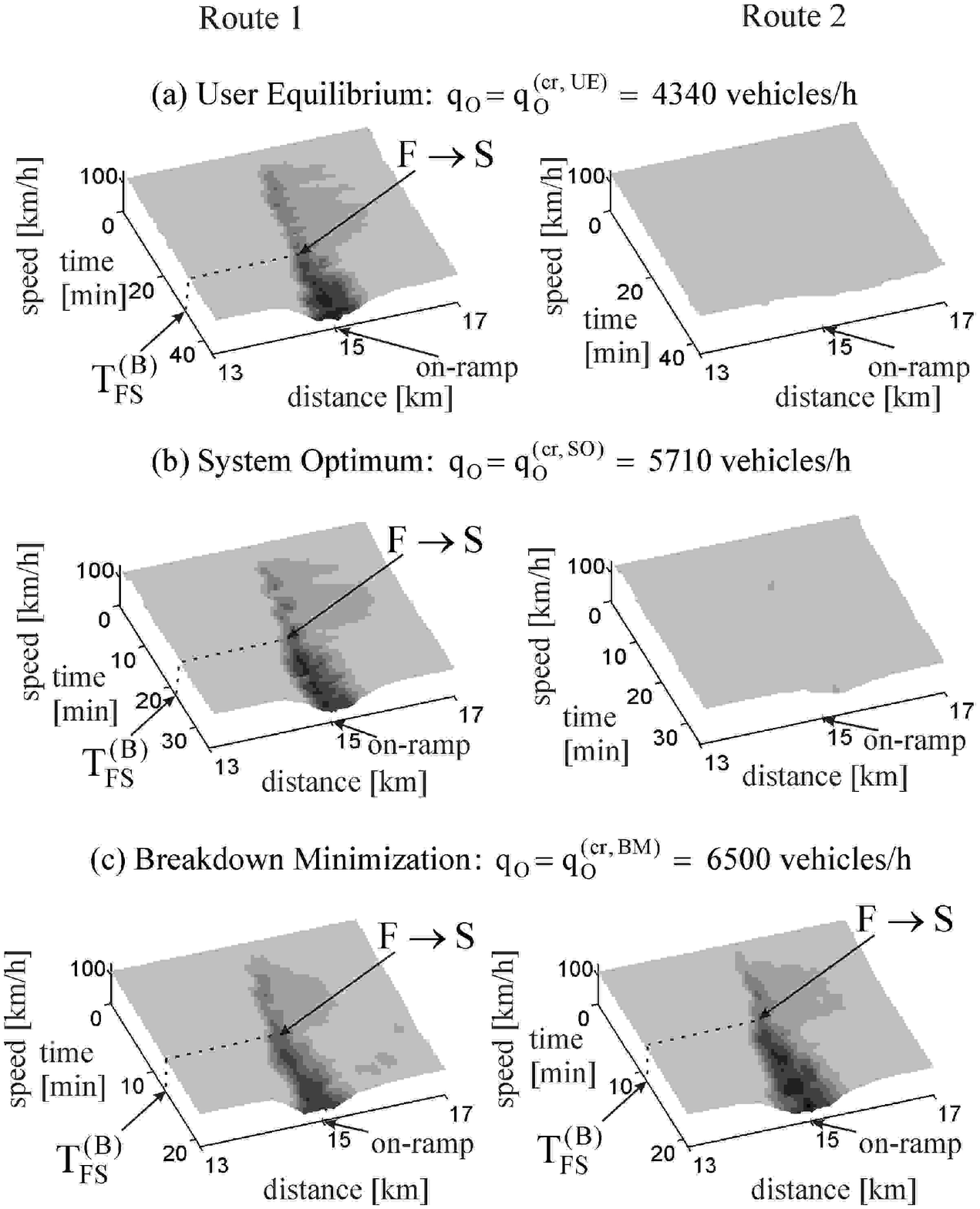}
\caption{Traffic breakdown under application of UE (\ref{UE_Eq}) (a), SO (\ref{SO_Eq}) (b)
   and BM principles (\ref{3Phase2ex}) (c), respectively. Speed in time and space in the right lane
  on  route 1 (left) and route 2 (right). In (a) $q_{1}= 3250$, $q_{2}=1090 $, $q^{\rm (cr, \ UE)}_{\rm O}=$ 4340 vehicles/h.
  In (b) $q_{1}  =$ 3250, $q_{2}=$ 2460, 
  $q^{\rm (cr, \ SO)}_{\rm O}=$ 5710 vehicles/h. In (c) $q^{\rm (cr, \ BM)}_{\rm O}=$ 6500 vehicles/h.
     $T^{\rm (B)}_{\rm FS}$ is a  random time delay of traffic breakdown labeled by
  arrows F$\rightarrow$S:
    $T^{\rm (B)}_{\rm FS}=$ 30 (a),  21 (b), 13 (route 1) and 11 (route 2) (c) min.
    $q_{\rm on   1}=q_{\rm on   2}=$   1000  vehicles/h; road location of on-ramp bottleneck $x_{\rm on}=$ 15 km; $L_{1}=$ 20, $L_{2}=$ 25 km.
 }
\label{Comparison}  
\end{center}
\end{figure}

We find that  a critical flow rate $q_{\rm O}=q^{\rm (cr)}_{\rm O}$ for traffic breakdown at one of the network bottlenecks,
   i.e., the inflow rate $q_{\rm O}$ at which      the breakdown occurs with probability $P^{\rm (B)}_{\rm FS}=1$
  on  route 1 or/and  2 in the network (Fig.~\ref{Network_Model} (a))~\cite{Realization}, satisfies  conditions
   \begin{eqnarray}
 q^{\rm (cr, \ BM)}_{\rm O} > 
  q^{\rm (cr, \ SO)}_{\rm O} >
 q^{\rm (cr, \ UE)}_{\rm O},
  \label{Com_par_Eq}
 \end{eqnarray}
 where   superscripts BM,   UE, and SO  are related to (\ref{3Phase2ex}), (\ref{UE_Eq}), and (\ref{SO_Eq}),
   respectively.  
  
   Under application of Wardrop's  UE principle (\ref{UE_Eq}),   most vehicles move on the  route 1 because it is shorter, i.e., $q_{1}>q_{2}$.
 This 
  explains why traffic breakdown occurs on  route 1
  (Fig.~\ref{Comparison}(a)). At the same flow rate $q_{\rm O}=$ 4340 vehicles/h, under application
  of the BM principle (\ref{3Phase2ex}) we find $P^{\rm (B,{\it k})}_{\rm FS}=0$ for $k=$ 1 and 2, because for the BM principle (\ref{3Phase2ex})
  values $q_{1}+q_{\rm on1}$ and $q_{2}+q_{\rm on2}=$ 3170 vehicles/h
  are smaller than $q^{\rm (B)}_{\rm th}\approx $ 3760 vehicles/h  (i.e.,
   $q^{\rm (B)}_{\rm th}\approx $ 1880 vehicles/h/lane,  
    Fig.~\ref{Network_Model} (e)).
  
  As the UE principle (\ref{UE_Eq}), the SO principle (\ref{SO_Eq}) leads also to $q_{1}>q_{2}$;
  however,  
  the   difference $q_{1}-q_{2}$ is not   great; therefore, the critical flow rate increases (Fig.~\ref{Comparison} (b)).
   At the same flow rate $q_{\rm O}=$ 5710 vehicles/h, under application
  of the BM principle (\ref{3Phase2ex}) we find $P^{\rm (B,{\it k})}_{\rm FS}=$ 0.05 for $k=1,2$; however, even when
  traffic breakdown occurs,  the resulting congested patterns exists  only during    
    about 10 min dissolving later due to a return S$\rightarrow$F transition (simulations made are not shown here).
   
   The greatest critical flow rate $q^{\rm (cr, \ BM)}_{\rm O}=$ 6500 vehicles/h is found for the BM principle (\ref{3Phase2ex}); in this case,
   traffic breakdown occurs on both routes 1 and 2 (Fig.~\ref{Comparison} (c))~\cite{ApBM}.

\begin{figure}
\begin{center}
\includegraphics*[width=14 cm]{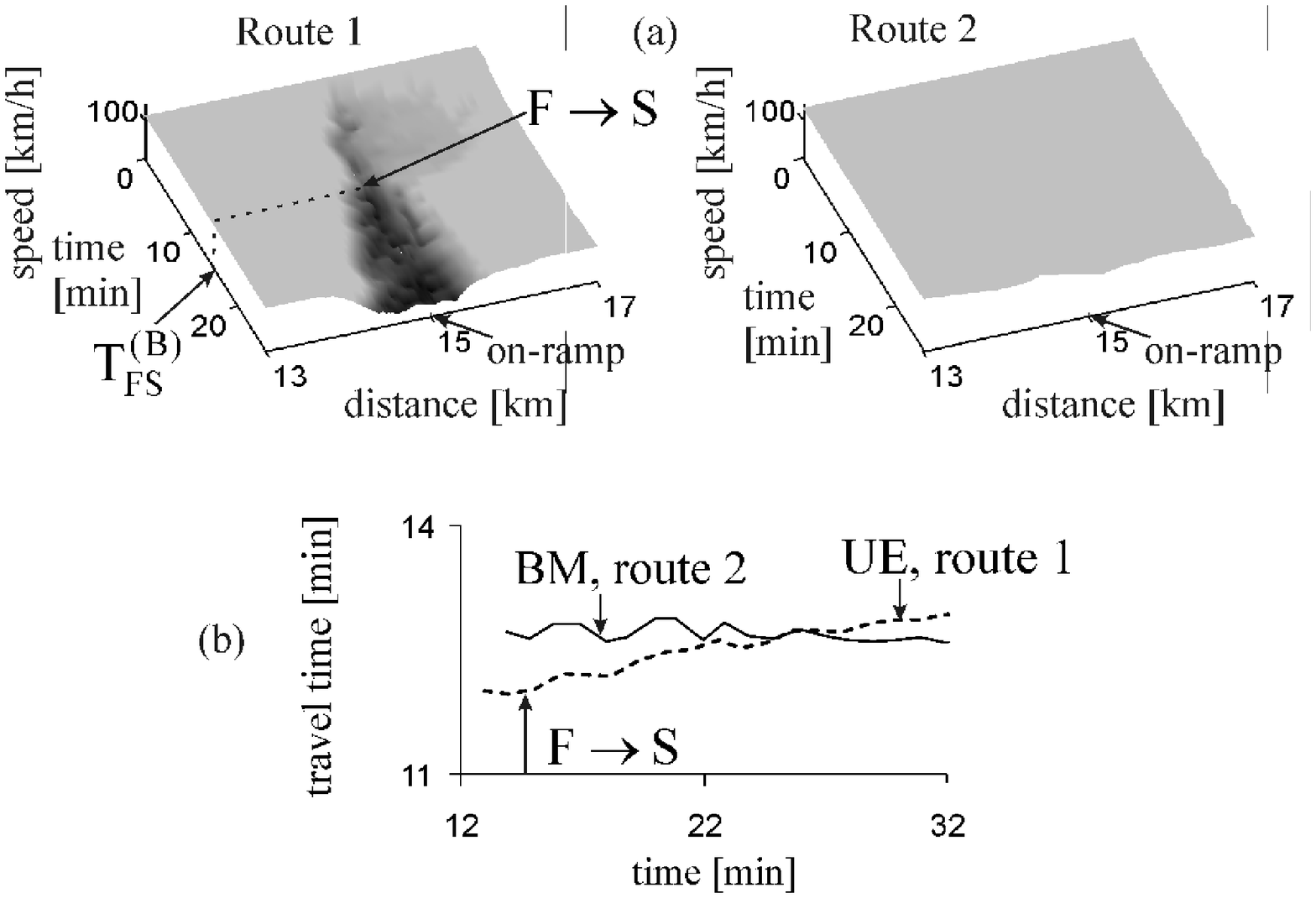}
\caption{Comparison of travel times under application of Wardrop's  UE  (\ref{UE_Eq}) and BM  (\ref{3Phase2ex}) principles at $q_{\rm O}=$ 4680 vehicles/h:  
  (a) Speed in the right lane in space and time for route 1 (left) and 2 (right) under application of   (\ref{UE_Eq})
  ($q_{1}=$ 3360, $q_{2}=$ 1320 vehicles/h).
  (b) Time-dependences of travel times on route 1 for  (\ref{UE_Eq}) (dashed curve)
  and on route 2 for  (\ref{3Phase2ex}) (solid curve).  
    $T^{\rm (B)}_{\rm FS}=$ 15 min. Under application of (\ref{3Phase2ex}), $P^{\rm (B,{\it k})}_{\rm FS}=0, \ k=1,2$.
    Other parameters are the same as those in Fig.~\ref{Comparison}. 
\label{Comparison2} } 
\end{center}
\end{figure}

   \begin{figure}
\begin{center}
\includegraphics*[width=14 cm]{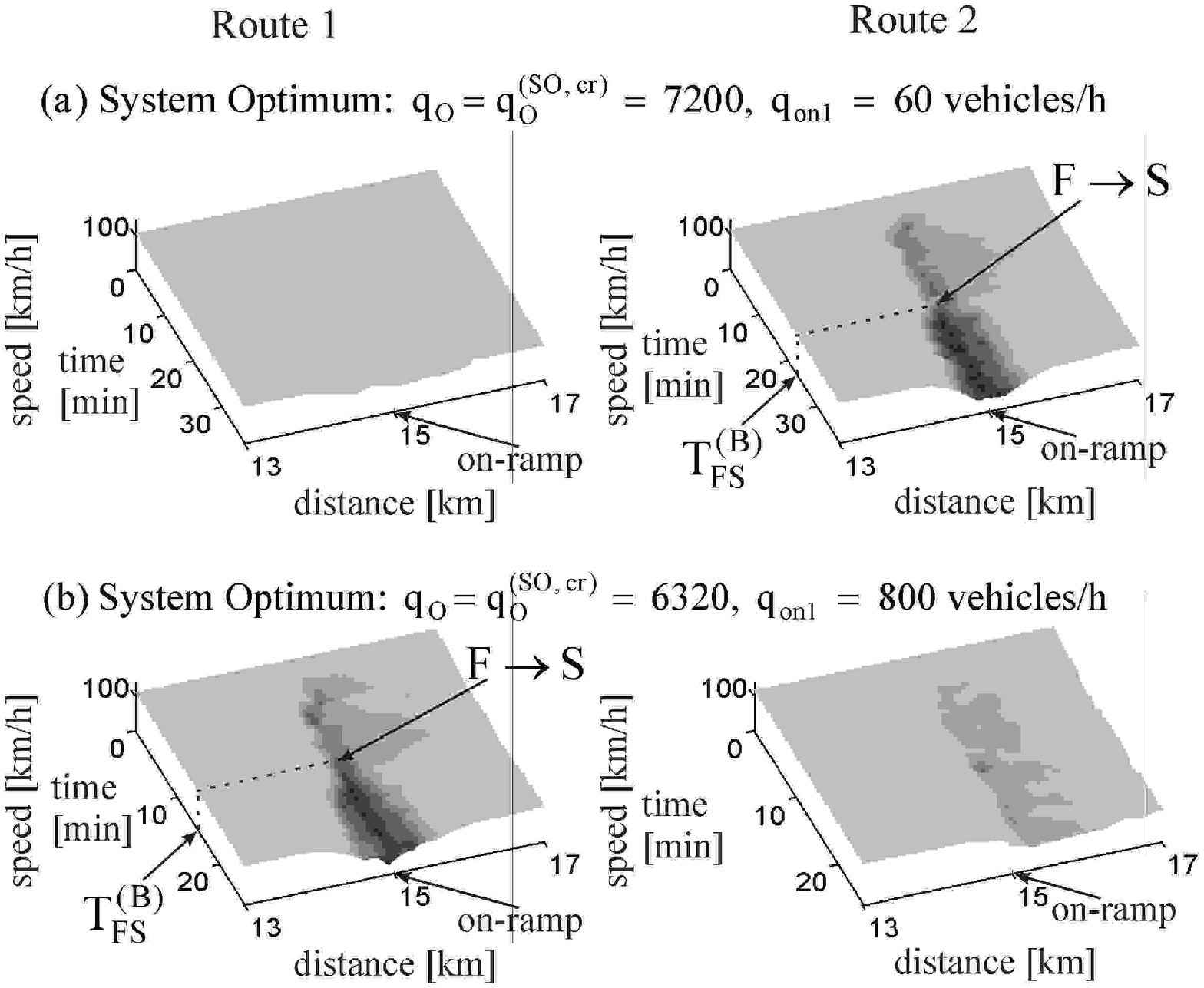}
\caption{Change in route on which breakdown occurs with  probability
$P^{\rm (B)}_{\rm FS}=1$ under asymmetric  bottleneck parameters and application of SO principle (\ref{SO_Eq}):
(a)   Traffic breakdown on route 2.
(b) Traffic breakdown on route 1. $q_{\rm on   2}=$ 1050 vehicles/h.
$(q_{\rm on   1}, \ q^{\rm (cr, \ SO)}_{\rm O})=$ (60, 7200) (a), (800, 6320) (b) vehicles/h. 
$q^{\rm (ch)}_{\rm on1}\approx$ 350 vehicles/h. $L_{2}=$ 23 km.
$T^{\rm (B)}_{\rm FS}=$ 22 (a) and 15 (b) min. Under application of     (\ref{3Phase2ex}),
  $P^{\rm (B,{\it k})}_{\rm FS}<1$ for $k=1,2$. Other parameters are the same as those in Fig.~\ref{Comparison}.
\label{SO_onramp} } 
\end{center}
\end{figure}

Thus in comparison with Wardrop's  UE and SO principles,  the advantage of the BM principle (\ref{3Phase2ex}) is the smaller traffic breakdown probability
at the same network inflow rate  and, therefore, the greater critical network inflow rate. The disadvantage of the BM principle (\ref{3Phase2ex})
is that more drivers move on  route 2 with a longer travel time. However, this disadvantage is true at small enough network inflow rates  only.
At greater network inflow rates,
 because of traffic congestion resulting from traffic breakdown   under application of the Wardrop's  UE and SO principles,
 we find a quick growth of travel time on the shorter route 1. 
 The greater network inflow rate exceeds the critical rate, the shorter the mean time delay of traffic breakdown and
the quicker the growth of   congestion.

For an example shown in Fig.~\ref{Comparison2}, under application of Wardrop's  UE principle (\ref{UE_Eq})  due to congestion on  route 1   
travel time on this route becomes as long as
under application of the BM principle (\ref{3Phase2ex})~\cite{Timelag}.

 Above 
 we have used symmetric bottleneck parameters 
 $q_{\rm on   1}= q_{\rm on   2}$  for which under application of 
 Wardrop's    principles  traffic breakdown occurs always on the shorter route 1  (Figs.~\ref{Comparison} and~\ref{Comparison2}). 
 Under asymmetric bottleneck parameters, we find the effect of change in route on which traffic breakdown can occur (Fig.~\ref{SO_onramp}):
  When $q_{\rm on   1}\ll q_{\rm on   2}$, traffic breakdown occurs on the longer route 2 (Fig.~\ref{SO_onramp}(a)),
 whereas at considerably greater flow rates  $q_{\rm on   1}$ traffic breakdown occurs on route 1 (Fig.~\ref{SO_onramp}(b)) as that     
 in Figs.~\ref{Comparison} and~\ref{Comparison2}.
 Thus for a given $q_{\rm on  2}$, there is a single flow rate $q_{\rm on   1}=q^{\rm (ch)}_{\rm on1}(q_{\rm on   2})$ for which
 $P^{\rm (B,1)}_{\rm FS}=P^{\rm (B,2)}_{\rm FS}$, i.e., $q^{\rm (cr, \ SO)}_{\rm O}  =
  q^{\rm (cr, \ BM)}_{\rm O}=$ 6900 vehicles/h; however, for all other flow rates  $q_{\rm on   1}$ condition (\ref{Com_par_Eq}) is valid.
 
   \section{BM principle and  
   traffic optimization   at single bottleneck \label{Bott_Net}}
   
 Breakdown probability at any single bottleneck   exhibits {\it no}
 minimum: the  breakdown probability is always a monotonously increasing flow rate function (Fig.~\ref{Network_Model} (e)).
 For this reason,   the minimization of breakdown probability $P^{\rm (B)}_{\rm FS}$ for a {\it single bottleneck}
 is not possible. However, the minimization of breakdown probability
 $P^{\rm (N)}_{\rm FS, net}$ (\ref{3Phase}) for 
 a {\it traffic network}  is 
  possible,   as   formulated in the BM principle
 of Sect.~\ref{S_BM}. 
 
  \begin{figure}
\begin{center}
\includegraphics*[width=11 cm]{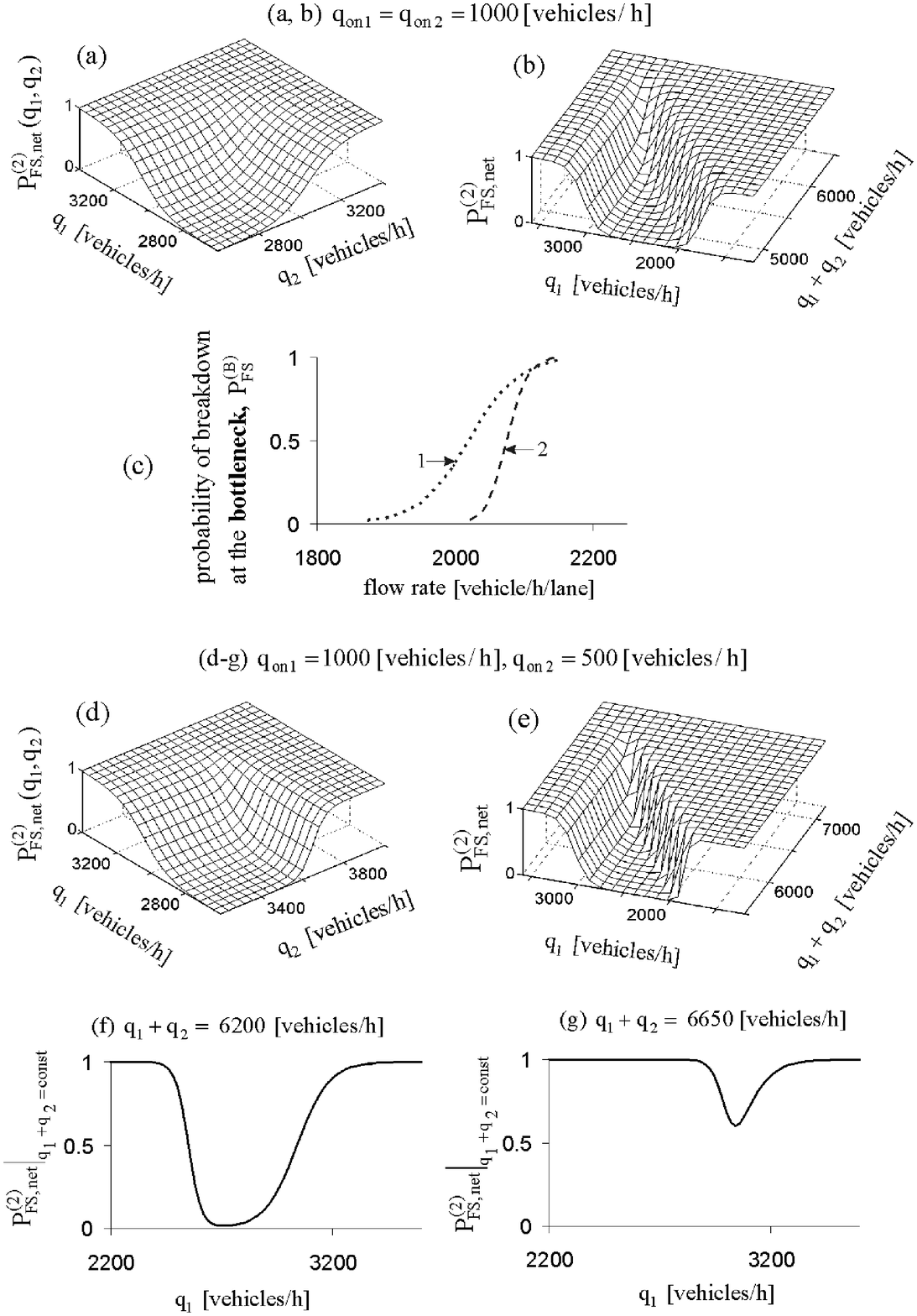}
\caption{Comparison of BM principle and  
   traffic optimization   at single bottleneck:  
  (a, b) Probability of traffic breakdown $P^{\rm (2)}_{\rm FS, net}$  
  in the {\bf network} with two bottlenecks shown in Fig.~\ref{Network_Model} (a)
 (i.e., when in Eq. (\ref{3Phase}) the value $N=2$) 
    as a function of the flow rates $q_{1}$ and $q_{2}$ (a)
   and a function of the flow rates $q_{1}$ and $q_{\rm O}=q_{1}+q_{2}$ (b) for   symmetric bottleneck parameters 
 $q_{\rm on   1}= q_{\rm on   2}=$    1000 [vehicles/h].
 (c)  Probability of traffic breakdown $P^{\rm (B)}_{\rm FS}$  at a {\bf single on-ramp  bottleneck}  
  as  a function of the flow rate downstream of the bottleneck for on-ramp inflow rates $q_{\rm on}=$ 1000 [vehicles/h] (curve 1
  that is the same as that in Fig.~\ref{Network_Model} (e))
  and $q_{\rm on}=$ 500 [vehicles/h] (curve 2).
  (d, e)      
$P^{\rm (2)}_{\rm FS, net}$  
    as   functions of  $q_{1}$ and $q_{2}$ (d)
   and   of   $q_{1}$ and $q_{\rm O}=q_{1}+q_{2}$ (e) for   asymmetric bottleneck parameters 
 $q_{\rm on   1}=  $    1000 and $ q_{\rm on   2}=$ 500 [vehicles/h].
 (f, g)  $P^{\rm (2)}_{\rm FS, net}$ as  a function of $q_{1}$     
      for   different  given values $q_{\rm O}=q_{1}+q_{2}=$ 6200 (f) and 6650 (g) [vehicles/h] associated with figure (e).
\label{Minimum} } 
\end{center}
\end{figure}

 To understand the sense of this conclusion, we consider
 the simple network shown in Fig.~\ref{Network_Model} (a). There are   {\it two} different bottlenecks
 in this case and, therefore, traffic assignment in the network
 changes  breakdown  probabilities for both bottlenecks.
 For this reason, although    breakdown probability for each of the bottlenecks separately
 has {\it no} minimum,
 there is a minimum in breakdown probability $P^{\rm (N)}_{\rm FS, net}$ (\ref{3Phase}) for 
 the network (Fig.~\ref{Minimum} (a, b)).

 Thus   the BM principle for the optimization of a traffic network is conceptionally 
  different in comparison with   known
  traffic optimization approaches  at a single bottleneck, in particular, with  on-ramp metering. 
  
   Figures~\ref{Minimum} (a, b) correspond to symmetric bottleneck parameters in the network shown  in Fig.~\ref{Network_Model} (a); this explains why the minimum
 of breakdown probability in the network  
 is related to the  condition  $q_{1}=q_{2}$ (Fig.~\ref{Minimum} (a)). However, breakdown probability at the on-ramp bottleneck
 $P^{\rm (B)}_{\rm FS}$ depends on the on-ramp inflow rate considerably (Fig.~\ref{Minimum} (c)). For this reason,
 under asymmetric bottleneck parameters the minimum
 of breakdown probability $P^{\rm (N)}_{\rm FS, net}$ in the network  shown  in Fig.~\ref{Network_Model} (a)
  is usually related to   condition  $q_{1}\neq q_{2}$ (Fig.~\ref{Minimum} (d, f, g)).

 \section{Conclusions}

\begin{description}
 \item {1.}  The network breakdown minimization (BM) principle    introduced in the article
  states
that the network optimum      is reached, when  link   flow rates   
are assigned  in the network in such a way that the  
probability for spontaneous occurrence of
 traffic breakdown at one of the network bottlenecks  during a given observation time
   reaches the minimum possible value;  this is equivalent to the maximization of   probability that 
traffic breakdown occurs at {\it none} of the network bottlenecks.
  We have shown  that the maximum network inflow rate at which free flows
 still remain in the   network   is considerably greater under
 application of  the BM   principle  
  than that under application  of the  Wardrop's   UE or
  SO principles. 
  \item {2.}  A traffic network optimization that is consistent with the  
  empirical features of traffic breakdown of Sect.~\ref{Int} can  consist of the stages:
  \begin{description}
 \item {(i)}  The minimization of traffic breakdown probability in the network based on
  the BM principle introduced in this article.
\item {(ii)}   A spatial limitation of   congestion growth, when  traffic 
breakdown has nevertheless occurred  at a network bottleneck, with the subsequent
 congestion dissolution at the bottleneck, if the dissolution of congestion due to traffic management
 in a neighborhood of the bottleneck is possible. An example  of this stage
  is the ANCONA on-ramp metering method~\cite{KernerBook,KernerBook2}.
   \end{description}
  
 A further development of this approach could be an interesting
 task for future investigations.
  \end{description}

\appendix

\section{Discrete Version of Kerner-Klenov Stochastic Three-Phase Traffic Flow Model and Model Parameters \label{CA_Ap}}

 A   traffic flow model used in this article (Tables~\ref{table_CA}--\ref{table2}) is a discrete version~\cite{KKl2009} of  the  
 Kerner-Klenov  stochastic three-phase traffic flow model of Ref.~\cite{KKl2003A}:
   rather than the continuum space co-ordinate,   
a discretized space co-ordinate with a small enough value  of the discretization cell $\delta x$  is used. Consequently,  
  the vehicle speed and acceleration (deceleration) discretization intervals are $\delta v$= $\delta x/\tau$
 and   $\delta a$= $\delta v/\tau$, respectively, where time step $\tau=$ 1 s. Because in 
  the discrete model version     discretized (and dimensionless) speed and acceleration 
 are used, which are measured respectively in the discretization values $\delta v$ and  $\delta a$,
 the value $\tau$ in all formulae below is assumed to be the dimensionless value $\tau=1$. 
 Explanations of the physics of vehicle motion   rules  in this model  can be found in Sect.~16.3 of~\cite{KernerBook}. 

 A choice of $\delta x$ in the discrete model
 version determines the accuracy of vehicle speed calculations {\it in comparison} with the initial continuum in space stochastic model  of~\cite{KKl2003A}.
 We have found that    the discrete   model   exhibits
similar characteristics of phase transitions and
resulting congested patterns at highway bottlenecks as those
in the
 continuum  model                     
 at $\delta x$  that satisfies the conditions
\begin{equation}
\delta x/\tau^{2} \ll b, \ a^{\rm (a)}, \ a^{\rm (b)}, \ a^{\rm (0)}.
\label{cond_Stoch}
\end{equation}

\begin{table}
\caption{Discrete version of stochastic model}
\label{table_CA}
\begin{center}
\begin{tabular}{|l|}
\hline
\multicolumn{1}{|c|}{
$v_{ n+1}=\max(0, \min({v_{\rm free,n}, \tilde v_{ n+1}+\xi_{ n}, v_{ n}+a
\tau, v_{{\rm s},n} }))$,   
}\\
\multicolumn{1}{|c|}{
$x_{n+1}= x_{n}+v_{n+1}\tau$,
}\\
\multicolumn{1}{|c|}{
$\tilde v_{n+1}=\max(0, \min(v_{\rm free,n},  v_{{\rm s},n}, v_{{\rm c},n})),
$
}\\
\multicolumn{1}{|c|}{
$v_{{\rm c},n}=\left\{\begin{array}{ll}
v_{ n}+\Delta_{ n} &  \textrm{at $g_{n} \leq G_{ n}$,} \\
v_{ n}+a_{ n}\tau &  \textrm{at $g_{n}> G_{ n}$}, \\
\end{array} \right.$
} \\
\multicolumn{1}{|c|}{
$\Delta_{ n}=\max(-b_{ n}\tau, \min(a_{ n}\tau, \ v_{ \ell,n}-v_{ n})),$
} \\
\multicolumn{1}{|c|}{
$v_{{\rm free}, n}=v_{\rm free}(g_{n})$, 
 $g_{n}=x_{\ell, n}-x_{n}-d$,
} \\
$\tau=$ 1; $a$ and $d$  are constants;
the lower index $\ell$ \\
marks variables related to the preceding vehicle. \\
\hline
\end{tabular}
\end{center}
\end{table}
\vspace{1cm} 

\begin{table}
\caption{Functions in  model I: Stochastic time delay of acceleration and
deceleration}
\label{table_CA1}
\begin{center}
\begin{tabular}{|l|}
\hline
\multicolumn{1}{|c|}{$a_{n}=a  \Theta (P_{\rm 0}-r_{\rm 1})$, \
$b_{n}=a  \Theta (P_{\rm 1}-r_{\rm 1})$,} \\
\multicolumn{1}{|c|}{
$P_{\rm 0}=\left\{
\begin{array}{ll}
p_{\rm 0} & \textrm{if $S_{ n} \neq 1$} \\
1 &  \textrm{if $S_{ n}= 1$},
\end{array} \right.
\quad
P_{\rm 1}=\left\{
\begin{array}{ll}
p_{\rm 1} & \textrm{if $S_{ n}\neq -1$} \\
p_{\rm 2} &  \textrm{if $S_{ n}= -1$},
\end{array} \right.$
}\\
\multicolumn{1}{|c|}{
$S_{ n+1}=\left\{
\begin{array}{ll}
-1 &  \textrm{if $\tilde v_{ n+1}< v_{ n}$} \\
1 &  \textrm{if $\tilde v_{ n+1}> v_{ n}$} \\
0 &  \textrm{if $\tilde v_{ n+1}= v_{ n}$},
\end{array} \right.$
}\\
$r_{1}={\rm rand}(0,1)$, $\Theta (z) =0$ at $z<0$ and $\Theta (z) =1$ at $z\geq 0$,
$p_{\rm 0}=p_{\rm 0}(v_{n})$, \\ $p_{\rm 2}=p_{\rm 2}(v_{n})$,
 $p_{\rm 1}$ is constant. \\
 \hline
\end{tabular}
\end{center}
\end{table}
\vspace{1cm}

\begin{table}
\caption{Functions  model II: Model speed fluctuations}
\label{table_CA2}
\begin{center}
\begin{tabular}{|l|}
\hline
\multicolumn{1}{|c|}{
$\xi_{ n}=\left\{
\begin{array}{ll}
\xi_{\rm a} &  \textrm{if  $S_{ n+1}=1$} \\
- \xi_{\rm b} &  \textrm{if $S_{ n+1}=-1$} \\
\xi^{(0)} &  \textrm{if  $S_{ n+1}=0$},
\end{array} \right.$
}\\
\multicolumn{1}{|c|}{$\xi_{\rm a}=a^{(\rm a)} \tau \Theta (p_{\rm a}-r)$, \
$\xi_{\rm b}=a^{(\rm b)} \tau \Theta (p_{\rm b}-r)$,} \\
\multicolumn{1}{|c|}{
$\xi^{(0)}=a^{(0)}\tau \left\{
\begin{array}{ll}
-1 &  \textrm{if $r\leq p^{(0)}$} \\
1 &  \textrm{if $p^{(0)}< r \leq 2p^{(0)}$ and $v_{n}>0$} \\
0 &  \textrm{otherwise},
\end{array} \right.$
}\\
$r={\rm rand}(0,1)$;
$a^{(\rm a)}=a^{(\rm a)} (v_{n})$, $a^{(\rm b)}=a^{(\rm b)} (v_{n})$;  \\
$p_{\rm a}$, $p_{\rm b}$, $p^{(0)}$, 
 $a^{(0)}$ 
are constants.\\
\hline
\end{tabular}
\end{center}
\end{table}
\vspace{1cm} 

\begin{table}
\caption{Functions in   model III: Synchronization gap $G_{n}$}
\label{table_CA3}
\begin{center}
\begin{tabular}{|l|}
\hline
\multicolumn{1}{|c|}{
$G_{n}=G(v_{n}, v_{\ell,n})$,
} \\
\multicolumn{1}{|c|}{
$G(u, w)=\max(0,  \lfloor k\tau u+  a^{-1}\phi_{0}u(u-w) \rfloor),$
} \\
$k$  ($k>1$) and $\phi_{0}$ are constants, \\ $\lfloor z \rfloor$ denotes the  integer part of a real number $z$. \\
\hline
\end{tabular}
\end{center}
\end{table}
\vspace{1cm} 

\begin{table}
\caption{Functions in   model IV: Safe speed $v_{{\rm s},n}$}
\label{table_CA4}
\begin{center}
\begin{tabular}{|l|}
\hline
\multicolumn{1}{|c|}{
$v_{{\rm s},n}=
\min{(v^{\rm (safe)}_{ n},  g_{ n}/ \tau+ v^{\rm (a)}_{ \ell})},$
} \\
\multicolumn{1}{|c|}{
$v^{\rm (a)}_{\ell}=
\max(0, \min(v^{\rm (safe)}_{ \ell, n}, v_{ \ell,n}, g_{ \ell, n}/\tau)-a\tau),$
} \\
\multicolumn{1}{|c|}{
$v^{\rm (safe)}_{ n}=\lfloor v^{\rm (safe)} (g_{n}, \ v_{ \ell,n}) \rfloor$ 
} \\
is        taken  as that in~\cite{Kra}, 
 which is a solution of \\ the
 Gipps's equation~\cite{Gipps} \\
 \multicolumn{1}{|c|}{
$v^{\rm (safe)} \tau_{\rm safe} + X_{\rm d}(v^{\rm (safe)}) = g_{n}+X_{\rm d}(v_{\ell, n})$,
} \\
where   $\tau_{\rm safe}$
 is a safe time gap, \\
 \multicolumn{1}{|c|}{
$X_{\rm d} (u)=b \tau^{2} \bigg(\alpha \beta+\frac{\alpha(\alpha-1)}{2}\bigg)$,
} \\
\multicolumn{1}{|c|}{
$\alpha=\lfloor u/b\tau \rfloor$ and $\beta=u/b\tau-\alpha$ 
} \\
are the integer and  fractional parts  of $u/b\tau$, \\
respectively; 
$b$ is constant. \\
\hline
\end{tabular}
\end{center}
\end{table}
\vspace{1cm}

  \begin{table}
\caption{Lane changing  occurring with probability $p_{\rm c}$ 
from the right lane to the left lane ($R \rightarrow L$)
and from the left lane to the right lane ($L\rightarrow R$)  and safety conditions
for lane changing~\cite{KKl2003A}
}
\label{table_lane}
\begin{center}
\begin{tabular}{|l|}
\hline
\multicolumn{1}{|c|}{Incentive conditions for lane  changing:} \\
\hline
\multicolumn{1}{|c|}{
$R \rightarrow L$: $v^{+}_{n} \geq v_{\ell, n}+\delta_{1}$   and $v_{n}\geq v_{\ell, n}$,
}\\
\multicolumn{1}{|c|}{
$L \rightarrow R$: $v^{+}_{n} > v_{\ell, n}+\delta_{1}$ or $v^{+}_{n}>v_{n}+\delta_{1}$.
}\\
In  conditions $R \rightarrow L$ and $L \rightarrow R$,   
the value $v^{+}_{n}$ at $g^{+}_{n}>L_{\rm a}$  \\ 
and the value  $v_{\ell, n}$ at $g_{n}>L_{\rm a}$ are replaced by $\infty$,  
where $L_{\rm a}$ is constant. \\
\hline
\multicolumn{1}{|c|}{Safety conditions for lane  changing:} \\ 
\hline
\multicolumn{1}{|c|}{
rules ($\ast $): \ $g^{+}_{n} >\min(v_{n}\tau, \ G^{+}_{n})$, \ $g^{-}_{n} >\min(v^{-}_{n}\tau, \ G^{-}_{n})$, \ where
}\\
\multicolumn{1}{|c|}{
$G^{+}_{n}=G( v_{n}, v^{+}_{n})$,
 $G^{-}_{n}=G(v^{-}_{n}, v_{n})$, 
 }\\ 
 \multicolumn{1}{|c|}{{\it or}   } \\
 \multicolumn{1}{|c|}{
rule ($\ast\ast $): \ $x^{+}_{n}-x^{-}_{n}-d > g^{\rm (min)}_{\rm target}$ \ with \ $g^{\rm (min)}_{\rm target}=\lfloor \lambda  v^{+}_{n} +d \rfloor$,
 }\\
 the vehicle should pass the midpoint point   \\
 \multicolumn{1}{|c|}{
$x^{\rm (m)}_{n}=\lfloor (x^{+}_{n}+x^{-}_{n})/2 \rfloor$
}\\
between   two neighboring vehicles in the target lane, i.e., \\
\multicolumn{1}{|c|}{
$\begin{array}{ll}
x_{n-1}< x^{\rm (m)}_{n-1} \  \textrm{and} \
 x_{n} \geq x^{\rm (m)}_{n} \\
 \textrm{or} \\
x_{n-1} \geq x^{\rm (m)}_{n-1} \  \textrm{and} \
 x_{n} < x^{\rm (m)}_{n}.
\end{array}$
}\\
 \hline 
 \multicolumn{1}{|c|}{Speed after lane changing:} \\
 \hline
\multicolumn{1}{|c|}{ $v_{ n}=\hat v_{n}$, \ $\hat v_{n}=  \min( v^{ +}_{n},  \ v_{n}+\Delta v^{(1)})$,
}\\ 
in $\hat v_{n}$ the speed $v_{n}$ is  related to    the initial lane before lane changing. \\
\hline
 \multicolumn{1}{|c|}{Vehicle coordinate after lane changing:} \\
 \hline
 Vehicle coordinate does not changes under the rules ($\ast $) \\
  and it   changes to $x_{n}=x^{\rm (m)}_{n}$  
under the rule ($\ast\ast $). \\
  $\lambda$, $\delta_{1}$, $\Delta v^{(1)}$ are constants; superscripts $+$  and  $-$  in variables, parameters, \\  and functions 
denote the preceding vehicle and the trailing vehicle \\
in the $\lq\lq$target"  (neighbouring)   lane, respectively; \\  the target lane is the 
lane into which the vehicle wants to change. \\  $G(u, w)$
is given in Table~\ref{table_CA3}. \\
\hline
\end{tabular}
\end{center}
\end{table}
\vspace{1cm}
 
 \begin{table}
\caption{Models of vehicle merging at on-ramp bottlenecks
that occurs when   a safety rule ($\ast$) {\it or} a safety rule  ($\ast \ast$) is satisfied~\cite{KKl2003A}
}
\label{table1}
\begin{center}
\begin{tabular}{|l|}
\hline
\multicolumn{1}{|c|}{Safety rule ($\ast$):}\\
\multicolumn{1}{|c|}{
$\begin{array}{ll}
g^{+}_{n} >\min(\hat  v_{n}\tau , \ G(\hat  v_{n}, v^{+}_{n})), 
g^{-}_{n} >\min(v^{-}_{n}\tau, \ G(v^{-}_{n},\hat  v_{n})),
\end{array} $
}\\
\multicolumn{1}{|c|}{
$\hat v_{n}=\min(v^{+}_{n},  \ v_{n}+\Delta v^{(1)}_{r}),$
} \\
in $\hat v_{n}$ the speed $v_{n}$ is  related to    the initial lane before lane changing, \\ $\Delta v^{(1)}_{r}>0$ is constant.\\
\hline
\multicolumn{1}{|c|}{Safety rule ($\ast \ast$):}\\
\multicolumn{1}{|c|}{
$x^{+}_{n}-x^{-}_{n}-d > \lfloor  \lambda_{\rm b} v^{+}_{n} +d \rfloor,$
}\\
\multicolumn{1}{|c|}{
$\begin{array}{ll}
x_{n-1}< x^{\rm (m)}_{n-1} \  \textrm{and} \
 x_{n} \geq x^{\rm (m)}_{n} \\
\ \textrm{or} \\
x_{n-1} \geq x^{\rm (m)}_{n-1} \  \textrm{and} \
 x_{n} < x^{\rm (m)}_{n},
\end{array}$
}\\
$\lambda_{\rm b}$ is constant. \\
\hline
\multicolumn{1}{|c|}{Parameters after vehicle merging:}\\
\multicolumn{1}{|c|}{$v_{n}=\hat v_{n}.$}\\
\multicolumn{1}{|c|}{Under the rule ($\ast $): $x_{n}$  maintains the
same,}\\
\multicolumn{1}{|c|}{under the rule ($\ast \ast$): $x_{n} = x^{\rm
(m)}_{n}$.}\\
\hline
\multicolumn{1}{|c|}{Speed adaptation before vehicle merging}\\
\multicolumn{1}{|c|}{
$v_{{\rm c},n}=\left\{\begin{array}{ll}
v_{ n}+\Delta^{+}_{ n} &  \textrm{at $g^{+}_{n} \leq G(v_{n}, \hat
v^{+}_{n})$,} \\
v_{ n}+a_{ n}\tau &  \textrm{at $g^{+}_{n}>G( v_{n}, \hat
v^{+}_{n})$}, \\
\end{array}\right. $
}\\
\multicolumn{1}{|c|}{
$\Delta^{+}_{ n}=\max(-b_{ n}\tau, \min(a_{ n}\tau, \ \hat v^{+}_{n}-v_{
n})),$
}\\
\multicolumn{1}{|c|}{
$\hat v^{+}_{n}=\max(0, \min(v_{\rm free,n}, \  v^{+}_{n}+\Delta
v^{(2)}_{r})),$
}\\
$\Delta v^{(2)}_{r}$ is  constant. \\
\hline
\end{tabular}
\end{center}
\end{table}
\vspace{1cm}

\begin{table}
\caption{Model parameters used in   simulations }
\label{table2}
\begin{center}
\begin{tabular}{|l|}
\hline
\multicolumn{1}{|c|}{Vehicle motion in road lane:
}\\
\hline
$\tau_{\rm safe}   = \tau=$ 1, $d = 7.5 \  \rm m/\delta x$, $\delta x=$ 0.01 m, \\
$v_{\rm free}(g)=\max [v_{\rm free}^{\rm (max)}(1-\kappa d/(g+d)), v_{\rm free}^{\rm (min)}]$, \\
$v^{\rm (max)}_{\rm free} \approx 38.9 \ {\rm ms^{-1}}/\delta v$ ($v^{\rm (max)}_{\rm free}=$ 140 km/h), \\

$v_{\rm free}^{\rm (min)} \ {\rm ms^{-1}}/\delta v$   is   constant      
  found from the system of equations: \\
$v^{\rm (free)}_{\rm min}= g^{\rm (free)}_{\rm min}/\tau$ and
$v^{\rm (free)}_{\rm min}=v_{\rm free}(g^{\rm (free)}_{\rm min})$ \\ ($v^{\rm (min)}_{\rm free}\approx$ 70 km/h),  
$b = 1 \ {\rm ms^{-2}}/\delta a$, $\delta v= 0.01 \  {\rm ms^{-1}}$,
 \\ $\delta a= 0.01 \  {\rm ms^{-2}}$,
$k=$ 3, $p_{1}=$ 0.3, $\phi_{0}=1$,
 $p_{b}=   0.1$, \\
$p^{(0)}= 0.005$, 
$p_{\rm 2}(v_{n})=0.48+ 0.32\Theta{( v_{n}-v_{21})}$, \\
$p_{\rm 0}(v_{n})=0.575+ 0.125\min{(1, v_{n}/v_{01})}$, \\
 $a^{(\rm b)}(v_{n})=0.2a+$ \\
  $+0.8a\max(0, \min(1, (v_{22}-v_{n})/\Delta v_{22})$, \\
  $a^{(0)}= 0.2a$, $\kappa=1.8$, 
  $a^{(\rm a)}= 0$, \\  
  $v_{22} = 12.5 \ {\rm ms^{-1}}/\delta v$,  
  $\Delta v_{22} = 2.778 \ {\rm ms^{-1}}/\delta v$, \\
$v_{01} = 10 \ {\rm ms^{-1}}/\delta v$, $v_{21} = 15 \ {\rm ms^{-1}}/\delta v$, $a=$ 0.5 ${\rm ms^{-2}}/\delta a$. \\
\hline
\multicolumn{1}{|c|}{Lane changing:
}\\
\hline  
$\delta_{1}=1$  
  ${\rm ms^{-1}}/\delta v$,  
   $L_{\rm a}=150 \  {\rm m}/\delta x$, \\
  $p_{\rm c}=0.2 $, 
 $\lambda=0.75$, 
   $\Delta v^{(1)}=2$
   ${\rm ms^{-1}}/\delta v$. \\  
    \hline   
   \multicolumn{1}{|c|}{On-ramp bottleneck model  (see   Fig.~16.2 of the book~\cite{KernerBook}):
}\\
 \hline
$\lambda_{\rm b}=$ 0.75,   
   $v_{\rm free \ on}=22.2 \ {\rm ms^{-1}}/\delta v$,  \\
   $\Delta v^{\rm (2)}_{\rm r}=$ 5   \ ${\rm ms^{-1}}/\delta v$  \\
   $L_{\rm r}=1 \ {\rm km}/\delta x$,   $\Delta v^{\rm (1)}_{\rm r}=10 \ {\rm ms^{-1}}/\delta v$,  \\
   $L_{\rm m}=$ 0.3 \     ${\rm km}/\delta x$. \\
   \hline
\end{tabular}
\end{center}
\end{table}


\begin{thebibliography}{8.}
\bibitem{FlowMaxMin}
The   flow rates $q^{\rm (B)}_{\rm th}$ and $q^{\rm (free \  B)}_{\rm max}$
 can depend  considerably on bottleneck characteristics and
 traffic parameters like the percentage of long vehicles, weather, etc.
In empirical observations,     $q^{\rm (free \  B)}_{\rm max}-q^{\rm (B)}_{\rm th}\approx$
600 $\div$ 1600 vehicles/h/lane~\cite{KernerBook,KernerBook2}.
\bibitem {TOB} 
  
  When in empirical observations of traffic breakdown the averaging time interval for traffic variables $T_{\rm av}$
  is longer than a characteristic time of the breakdown development (about 1 min), then
  one can apply $T_{\rm ob}=T_{\rm av}$ in the definition of breakdown probability (see explanations in Sec.~10.3.1 of the book~\cite{KernerBook}).
\bibitem{NM_B}
 Even if a network link is a spatially homogeneous road (i.e., the road without on- and off-ramps,
without road gradients, curves, etc.), nevertheless within a flow rate range
$q_{\rm th}\leq q \leq q^{\rm (free)}_{\rm max}$ traffic breakdown  occurs 
on this link spontaneously with   probability $P_{\rm FS}>$ 0, however, at a random location
within the link~\cite{KernerBook}. In (\ref{3Phase}), this link can also be considered containing a
  bottleneck with the threshold flow rate $q^{\rm (B)}_{\rm th}=q_{\rm th}$ 
  and maximum flow rate $q^{\rm (free \  B)}_{\rm max}=q^{\rm (free)}_{\rm max}$.
\bibitem{Timelag}
  There is 
 a time lag   about travel time  $T_{1}$ of a probe vehicle between the time instant of traffic breakdown  
and the time instant at which    
  flow rates $q_{1}$ and $q_{2}$ begin to change through the application of  (\ref{UE_Eq}) or   (\ref{SO_Eq})    under congestion conditions.
\bibitem{GM_model}
 The application of a three-phase
  traffic flow model is explained as follows.
  As shown in Sect.~10.3 of~\cite{KernerBook},
the first-order phase transition  in traffic flow models that belong to
the General Motors (GM) model class reviewed in~\cite{Helbing}
 is associated with an F$\rightarrow$J transition. In contrast,
 in empirical observations
 rather than the F$\rightarrow$J transition, an F$\rightarrow$S transition governs
 traffic breakdown at a bottleneck in vehicular traffic.
\bibitem{Realization} 
To find probability of traffic breakdown
 $P^{\rm (B)}_{\rm FS}$, a study of traffic breakdown in the 
 network
 is repeated for 40 different realizations for each given   flow rate.
 In these 40  
 realizations model parameters are the same, however, the initial conditions
 for random model fluctuations   are different.
\bibitem{ApBM}
 It should be noted that the application of the
  BM principle (\ref{3Phase2ex}) has a sense for free flows   only; therefore, the development
  of congested patterns after traffic breakdown has occurred is shown  in Fig.~\ref{Comparison} (c)   
  only  for the illustration of emergent congestion under unchanged   model parameters.
\end{thebibliography}
\end{document}